\documentclass[jcp,twocolumn,superscriptaddress,floatfix,graphicx,showpacs]{revtex4-1}
\usepackage{ifpdf}
 \newif\ifpdf
\ifx\pdfoutput\undefined
   \pdffalse
\else
   \pdfoutput=1
   \pdftrue
\fi
\ifpdf
   \usepackage{graphicx}
   \usepackage{epstopdf}
\else
   \usepackage{graphicx}
\fi

\usepackage{amsmath,amssymb}
\usepackage{color}

\usepackage{epsfig}
\usepackage{wasysym}
\usepackage{srctex}\usepackage{hyperref}%\usepackage[hypertex]{hyperref}

\DeclareMathOperator{\Real}{Re} \DeclareMathOperator{\Imag}{Im}
\DeclareMathOperator{\kb}{\mathit k_{\scriptscriptstyle B}}
\begin{document}

%\preprint{APS/123-QED}

%\title{Dynamical Correlation Functions in complex frequency plain. The singularity band of VAF$(\omega)$ in LJ fluid }
%\title{Complex-$\omega$ spectroscopy of Dynamical Correlation Function in classical particle system: the singularity band of velocity auto correlation function of LJ-fluid }
\title{Singularity band of velocity auto correlation function of Lennard-Jones fluid in complex $\omega$-plain }

\author{N.M. Chtchelkatchev}
\affiliation{Moscow Institute of Physics and Technology, 141700 Moscow, Russia}
%\affiliation{Institute for High Pressure Physics, Russian Academy of Sciences, 142190 Troitsk, Russia}
\affiliation{L.D. Landau Institute for Theoretical Physics, Russian Academy of Sciences, 142432, Moscow Region, Chernogolovka, Russia}
\affiliation{Department of Physics and Astronomy, California State University Northridge, Northridge, CA 91330, USA}

\author{R.E. Ryltsev}
\affiliation{Institute of Metallurgy, Ural Division of Russian Academy of Sciences, 620016 Yekaterinburg, Russia}
\affiliation{L.D. Landau Institute for Theoretical Physics, Russian Academy of Sciences, 142432, Moscow Region, Chernogolovka, Russia}
%65.20.De	General theory of thermodynamic properties of liquids, including computer simulation
%36.40.Qv	Stability and fragmentation of clusters
%61.20.Ne	Structure of simple liquids

\begin{abstract}
It is well known from the quantum theory of strongly correlated systems that poles (or more subtle singularities) of dynamic correlation functions in complex plane usually correspond to the collective or localized modes. Here we address singularities of velocity autocorrelation function $Z$ in complex $\omega$-plain for the one-component particle system with isotropic pair potential. We have found that naive few poles picture fails to describe analytical structure of $Z(\omega)$ of Lennard-Jones particle system in complex plain. Instead of few isolated poles we see the singularity manifold of $Z(\omega)$ forming branch cuts that suggests Lennard-Jones velocity autocorrelation function is a multiple-valued function of complex frequency. The brunch cuts are separated from the real axis by the well-defined ``gap''. The gap edges extend approximately parallel to the real frequency axis. The singularity structure is very stable under increase of the temperature; we have found its trace at temperatures even several orders of magnitude higher than the melting point. Our working hypothesis that the branch cut origin is related to the ``interference'' in $Z$ of one-particle kinetics and collective hydrodynamic motion.
\end{abstract}

\maketitle
\section{Introduction}
Dynamic correlation functions (DCF) are one of the main tools that allow understanding nature of condensed matter particle systems \cite{Lovesey1986DynCorrBook,hansenMcDonald, Rickayzen2013GreenBook,Abrikosov2009AGDBook}. Fourier spectra of DCF keep the information about the spectrum and inverse lifetime of collective excitations, particle diffusion and most other key properties of the system.  As a rule, using the results of numerical simulations, like molecular dynamics, one can find spectrum of DCF on real (or sometimes on imaginary) axis in the frequency $\omega$-space.  However, interesting fitches of the DCF should be hidden at complex $\omega$. It is well known from the quantum theory of strongly correlated systems that poles (or more subtle singularities) of DCF in complex plane usually correspond to the collective or localized modes. Then, for example, the real part of the pole position in the $\omega$-plane produces the energy of the excitation while the imaginary part corresponds to the inverse life time \cite{Rickayzen2013GreenBook,Abrikosov2009AGDBook}. Interesting question what singularities of DCF$(\omega)$ for classical particle system one can find in the complex $\omega$-plain.

Here we consider the onecomponent particle system with isotropic Lennard-Jones (LJ) pair potential and focus mainly on the velocity autocorrelation function (VAF) $Z(t)$.  It is well known that in liquid phase $Z(t)$ is nonmonotonic at short time scales $ t \sim\tau_0$, where $\tau_0$ is the period of the particle motion in the effective potential well formed by the surrounding particles  (i.e. the inverse Einstein frequency)\cite{Rice1966JChemPhys,hansenMcDonald,boon1991MolHydroBook}. In the dilute gas phase and in the supercritical fluid far above melting and critical temperatures $Z(t)$ decays monotonically with time at time scale of the order of the relaxation time of the particle diffusion: $\tau=m D/\kb T$, where $D$ is the diffusion coefficient, $T$ is the temperature and $m$ is the particle mass~\cite{Rice1966JChemPhys,hansenMcDonald,boon1991MolHydroBook}.

%The change from monotonic to nonmonotonic decay is the continuous process~\cite{Rice1966JChemPhys,hansenMcDonald,Ryltsev2014JCP,Ryltsev2013PRE}.

% At hydrodynamic time scales, much larger than $\tau$,  $Z(t)$ shows the universal time-decay tails governed by hydrodynamic fluctuations, see Refs.~\cite{hansenMcDonald,Ryltsev2014JCP}.

 %For example, $Z(t)\sim t^{-3/2}$ and $Z(\omega)\sim\omega^{-1/2}$-singularity for $\omega\tau\ll 1$ in the simple liquid confined in three-dimensional volume with a soft core interaction potential at small distances.

There are many approximations of $Z(t)$ for simple particle systems. It is well established that for satisfactory approximation of $Z(t)$ one should take more than one relaxation time in the memory function or even the continuum of the relaxation times~\cite{gaskell1978,Levesque1973PhysRevA,gaskell1978JPhys,Krishnan2003JCP,hansenMcDonald,boon1991MolHydroBook}. But how then the manifold of relaxation times and Einstein frequencies look like? Here we search for answers to these questions.

As far as we know there is no universal explicit expression that produces $Z(t)$ equally well at small and hydrodynamic (large) time scales~\cite{hansenMcDonald,boon1991MolHydroBook}. On the other hand, low approximation accuracy do not allow reliable investigation of singularity manifolds in the complex $\omega$-plane. Therefore here we investigate $Z(\omega)$ numerically and develop the machinery for numerical analytical approximation.

We see the singularity manifold of $Z(\omega)$ forming branch cuts that suggests LJ VAF is a multiple-valued function of complex frequency. The brunch cuts are approximately parallel to the real frequency axis and separated from it by the well-defined ``gap''. The singularity manifold is stretched along the real axis. Going higher and higher with temperature the singularities more and more group and $Z(\omega)$ better and better agrees with the exponential memory function approximation of $Z(t)$~\cite{hansenMcDonald}. The singularity structure is very stable under increase of the temperature; we have found its trace at temperatures even several orders of magnitude higher than the melting point.

When $t>t_h$, where $t_h$ is some characteristic {transient time for hydrodynamic regime}, $Z(t)$ has nontrivial power law decaying tail $\propto t^{-3/2}$. In frequency space that causes nonanalyticity of $Z(\omega)\sim\omega^{-1/2}$ at small $\omega$~\cite{Levesque1974PRL}. The short-time $t<t_h$ behaviour of $Z(t)$ with satisfactory  accuracy can be simulated by a number of damped and over damped oscillators that formally produce exponential long-time time decay. In frequency representations these oscillators produce analytical function with a number of poles.  Binding ``analytical'' oscillators with ``nonanalytical'' hydrodynamics at $t\sim t_h$ produces, from our point of view, nonanalyticity in the Fourier transform of $Z(t)$. We do see that the time scale $t_h$ well corresponds the half-width $\Delta \omega$ of the {gap between branch cuts} in $Z(\omega)$ in wide temperature range: $t_h\approx 2\pi/\Delta \omega$.

\section{Model calculations}
One way to go into the complex $\omega$-plain is the $z$-transform of DCF$(t)$ [this is fast complex-$\omega$ Fourier transform]. This approach has been used in Refs.~\cite{kneller2001JCP,Krishnan2003JCP}. However the analytical continuation of DCF were not there the purpose of the study except the answer to the question if the singularities of the memory function belong to the stability manifold $|z|<1$ of the $z$-transform. Since the stability of $z$-transform is limited in the complex $\omega$-plain one should search for alternatives. Other traditional methods of the analytical continuation, like integration of Cauchy-Riemann equations or different methods of series reexpansion~\cite{reichel1986numerical}, also are unstable approaching DCF$(\omega)$ singularities.

The promising way to study the singularities of DCF in the complex $\omega $-plain is to built at real $\omega$ an approximation of DCF by a meromorphic function and finally do the analytical continuation of it. [In complex analysis, meromorphic function is a function that is holomorphic except a set of isolated points~\cite{abramowitz1964stegun,lavrentiev1973Shabat}.] The Pade-approximation is the keystone of one of the analytical continuation methods that follows this receipt~\cite{ferris1973numerical,EliashbergEqPade,baker1996pade,yamada2013analyticity}. Here we perform complex-$\omega$ spectroscopic investigation of $Z(\omega)$ based on  the Pade-approximation, while $Z(t)$ we find from Molecular Dynamic (MD) simulations. %We answer the question what part of the DCF-spectrum (characteristic time scales) at real $\omega$ mainly generates these singularities after the numerical analytical continuation into the complex $\omega$-plain.

For MD simulations of $Z(t)$, we have used $\rm{DL\_POLY}$  Molecular Simulation Package~\cite{dl_poly} developed at Daresbury Laboratory.
For simulations we use the LJ pair potential model in a wide range of parameters. For LJ liquid we apply the standard pair potential, $U(r)=4\varepsilon [(\sigma/r)^{12}-(\sigma/r)^{6}]$, where $\varepsilon$ – is the unit of energy, and $\sigma$ is the core diameter. In the remainder of this paper we use the dimensionless quantities: $\tilde r= r/\sigma$, $\tilde U = U/\varepsilon$, temperature $\tilde T = T/\varepsilon$, density $\tilde{\rho}\equiv N \sigma^{3}/V$, and time $\tilde t=t/[\sigma\sqrt{m/\varepsilon}]$, where $m$ and $V$ are the molecular mass and system volume correspondingly. As we will only use these reduced variables, we omit the tildes.
\begin{figure*}[tb]
  \centering
  \includegraphics[width=0.99\textwidth]{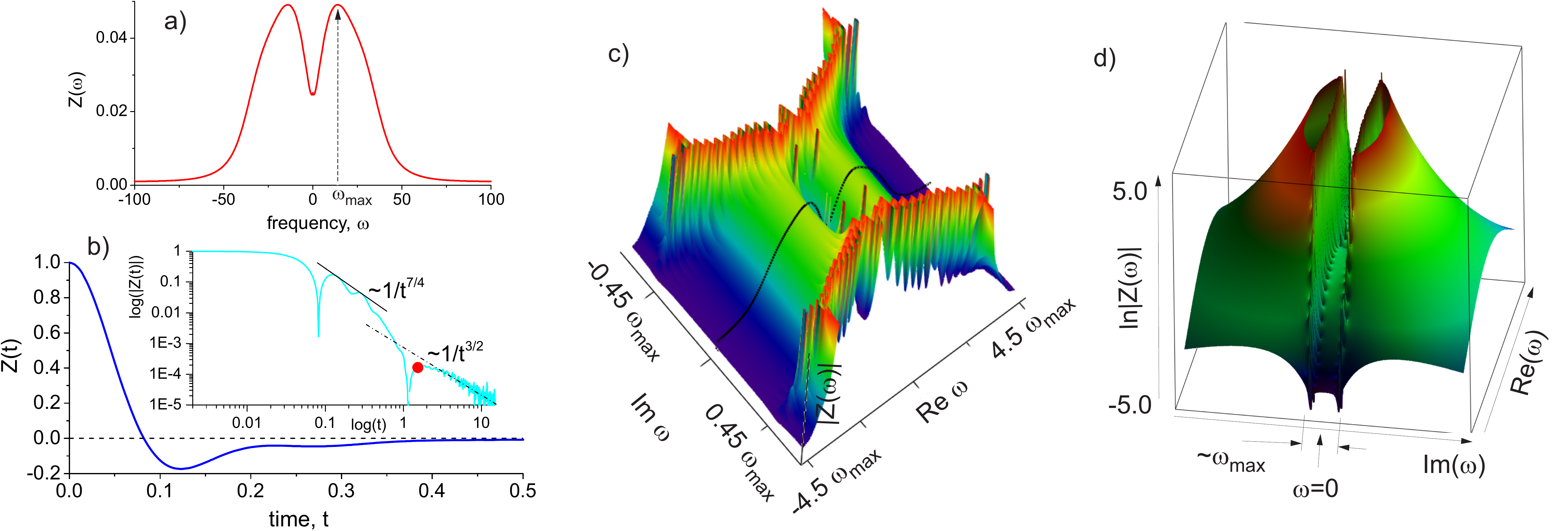}\\
  \caption{ VAF at real frequency axis (a) and in time representation (b) for $T=1.4$ and $\rho=1$: results of MD simulation.  The inset in (b) shows long time tail of $Z(t)$ in the double logarithmic scale; the red bullet \textcolor{red}{$\bullet$} points the time scale $\Delta t=2\pi/\Delta \omega$, where $\Delta \omega$ is the half-width of the branch cut. Graphs (c) and (d) show analytical continuation of $Z(\omega)$ into complex $\omega$-plain using multipoint Pade approximation built on top of $1400$ uniformly distributed knot-points in $(0,6.3\omega_{\rm max})$. 3D plot of $|Z(\omega)|$ in the complex $\omega$-plane; the black curve in (a) is $Z(\omega)$ at real $\omega$. Regular behaviour of $Z(\omega)$ for small imaginary frequencies ends abruptly by the ``walls'' of singularities. Graph (d) shows $\ln|Z(\omega)|$ in the complex frequency domain $\Real\omega,\,\Imag\omega\in(-4.5\omega_{\rm max},4.5\omega_{\rm max})$.}\label{fig1}
\end{figure*}

\begin{figure}[b]
  \centering
  \includegraphics[width=0.99\columnwidth]{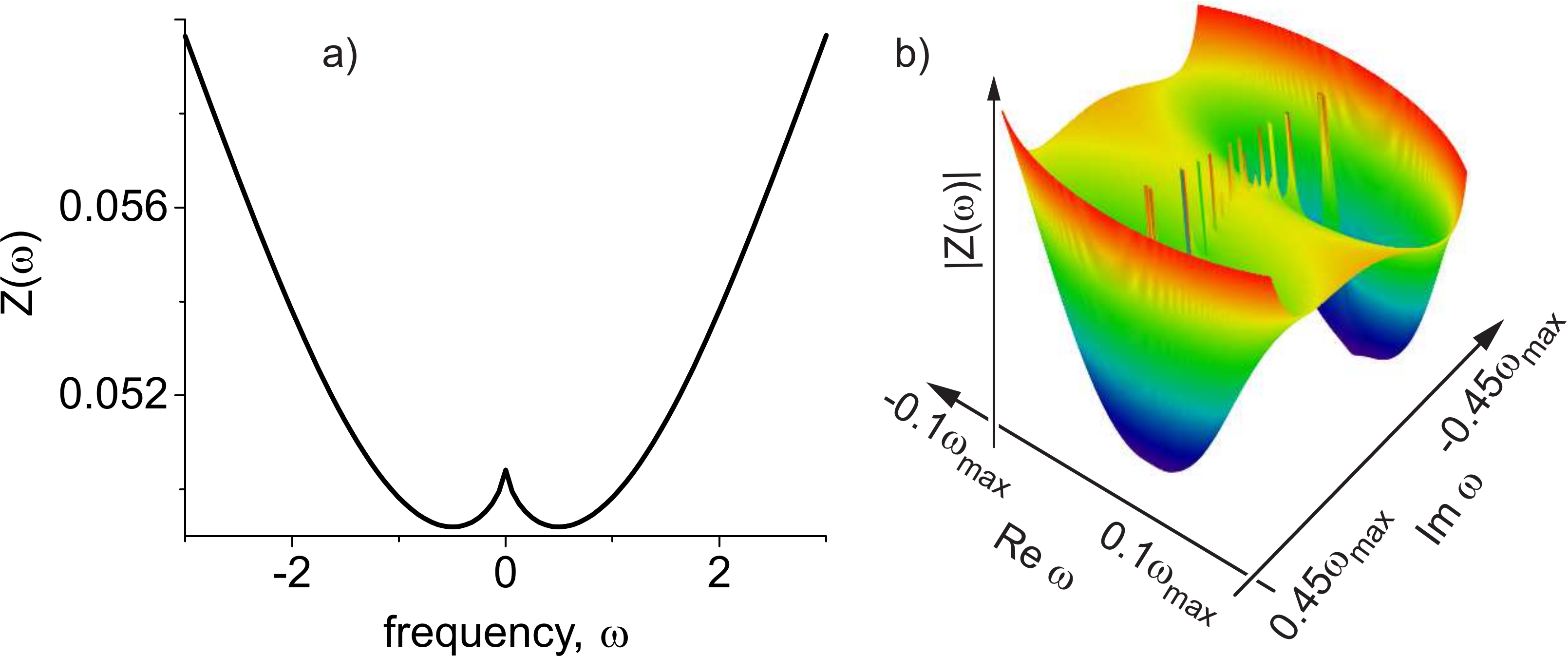}\\
  \caption{(a) Hydrodynamic tales of $Z(t)\sim t^{-3/2}$ generate additional singular terms in $Z(\omega\to 0)\sim {\rm const}-\sqrt{|\omega|}$ at real frequency axis. This nonanalyticity at $\omega=0$ produces the branch cut shown in (b) where we use the same Pade approximant as in Fig.~\ref{fig1}. }\label{fig2}
\end{figure}

For simulations, we have considered the system of $N=128^3\simeq 2.1\cdot 10^6$ particles that were simulated under periodic boundary conditions in 3-dimensional cube mostly in the Nose-Hover (NVT) and also in NPT ensambles. Such a large number of particles is necessary to correctly describe long time behaviour of correlation functions, see~Ref.~\cite{Ryltsev2014JCP}. We consider the system at fixed density $\rho=1$ and different temperatures in the interval $T\in(1.4,200)$. According to equilibrium temperature-density phase diagram~\cite{smit1992JCP,kalyuzhnyi1996,Frenkel2001book,lin2003JCP,Ryltsev2013PRE}, this range covers thermodynamic states from the liquid just above the melting line up to the supercritical fluid approaching the ideal gas limit.  The MD time step was $t=0.0001-0.001$ chosen so that to provide good energy conservation for given thermodynamic conditions.

To obtain perfect hydrodynamic tails of $Z(t)$ we improved the code of $\rm{DL\_POLY}$  Molecular Simulation Package~\cite{dl_poly} and inserted inside specially designed parallel MPI-code to calculate $Z(t)$ for very large systems~\cite{Ryltsev2014JCP}.  Calculations of $Z(t)$ tails for $2\cdot10^6$ particles requires at least 128 processors with $10-15~\rm{Gb}$ of operational memory per each one.
% Near the melting line ($\rho=1$, $T=1.4$) we needed at least $15~\rm{Gb}$ per processor.

\section{Results}

\subsection{Fluid just above the melting line}
\subsubsection{The branch cut}
We start our investigation from $T=1.4$ and $\rho=1$: these parameters correspond to fluid, just above the melting line. The results are shown in Figs.~\ref{fig1}~-~\ref{fig2}. $Z(t)$ obtained using MD simulation is shown in Fig.~\ref{fig1}(a) and Fig.~\ref{fig1}(b) represent $Z(\omega)$ obtained by Fourier transformation of $Z(t)$. At $t> 1$, $Z(t)$ becomes positive and as it should be in fluid~\cite{Williams2006PRL}, and at $t\gtrsim 5$ it demonstrates $t^{-3/2}$ asymptotic behaviour, see insert in Fig.~\ref{fig1}(b). We use this long time asymptotic to obtain good Fourier transform of $Z(t)$. We perform extrapolation of $Z(t)$ to long times by $at^{-3/2}$ asymptotic with the appropriate value of $a$. [Of course, we have tested that this procedure does not change $Z(\omega)$ behavior and does not influence the properties of analytical continuation to complex $\omega$-plain.] As the result we obtain smooth $Z(\omega)$ curve at all interesting values of $\omega$ (see Fig.~\ref{fig1}(b)). The $Z(\omega)$ curve has the form typical to that for simple liquids \cite{hansenMcDonald,boon1991MolHydroBook}. In particular, it demonstrates pronounced maximum at $\omega=\omega_{\rm max}$.

Figs.~\ref{fig1} (c) and (d) show analytical continuation of $Z(\omega)$ into complex $\omega$-plain using multipoint Pade approximation built on top of $1400$ uniformly distributed knot-points in $(0,6.3\omega_{\rm max})$. Technical aspects of Pade approximation can be found in Sec.~\ref{secMethods}. Building the Pade approximant we explicitly take into account that $Z(\omega)$ is even function at real $\omega$. So the continued fraction of Pade approximant is the function of $\omega^2$.

3D plot of $|Z(\omega)|$ in the complex $\omega$-plane is shown in Figs.~\ref{fig1} (c) and (d). Regular behaviour of $Z(\omega)$ for small imaginary frequencies ends abruptly by the ``walls'' of singularities constructed from poles (and zero nodes) of the Pade-approximant.  Fig.~\ref{fig1} (d) shows plot of $\ln|Z(\omega)|$ in the domain $\Real\omega,\,\Imag\omega\in(-4.5\omega_{\rm max},4.5\omega_{\rm max})$. Such series of poles and zeros is the way the Pade approximation typically represents brunch cuts of multi-valued functions (see Sec.~\ref {secMethods}).

In the insert in Fig.~\ref{fig1}(b) we show the long-time behaviour of $Z(t)$ in double logarithmic scales. The red bullet points the time scale $\Delta t=2\pi/\Delta \omega$, where, we remind, $\Delta \omega$ is the characteristic scale approximately equal to the half-width of the gap between branch cuts. As follows, $\Delta t\sim t_h$,where $t_h$ corresponds to crossover of system dynamics from kinetic to hydrodynamic regime.

\subsubsection{Hydrodynamic asymptotic of VAF: additional branch cut at small frequencies}
As we have already mentioned, at timescales much larger than inverse Einstein frequency VAF is positive and it has the following asymptotic behaviour: $Z(t)\sim t^{-3/2}$~\cite{hansenMcDonald,Ryltsev2014JCP}. It generates singular terms in $Z(\omega\to 0)\sim {\rm const}-\sqrt{|\omega|}$ at real frequency axis, see Fig.~\ref{fig2}a for illustration. This nonanalyticity at $\omega=0$ should produce additional branch cut. We do see it in Fig.~\ref{fig2}b where $Z(\omega)$ is shown. Preparing  Fig.~\ref{fig2}b we have used the same Pade approximant as we have used working on Fig.~\ref{fig1}(c)-(d). Hydrodynamic branch cut is on the imaginary axis and so it lies transversely to the branch cut presented in the Figs.~\ref{fig1}(c)(d). Note that for liquid near the melting line hydrodynamic singularity is located at only small vicinity of $\omega=0$ (compare Fig.~\ref{fig1}(a) and Fig.~\ref{fig2}(a)). Thus the hydrodynamic branch cut is only detectable at small frequency scales (compare Fig.~\ref{fig1}(c)(d) and Fig.~\ref{fig2}(d)).

\subsection{From fluid to gas: evolution of $Z(\omega)$ }
Below we test how stable is the the branch cut in $Z(\omega)$ when we increase the temperature of the fluid far above the melting line. It follows that the branch cut is very stable to temperature.

\subsubsection{$T=40$ and $\rho=1$}
\begin{figure}[t]
  \centering
  \includegraphics[width=0.99\columnwidth]{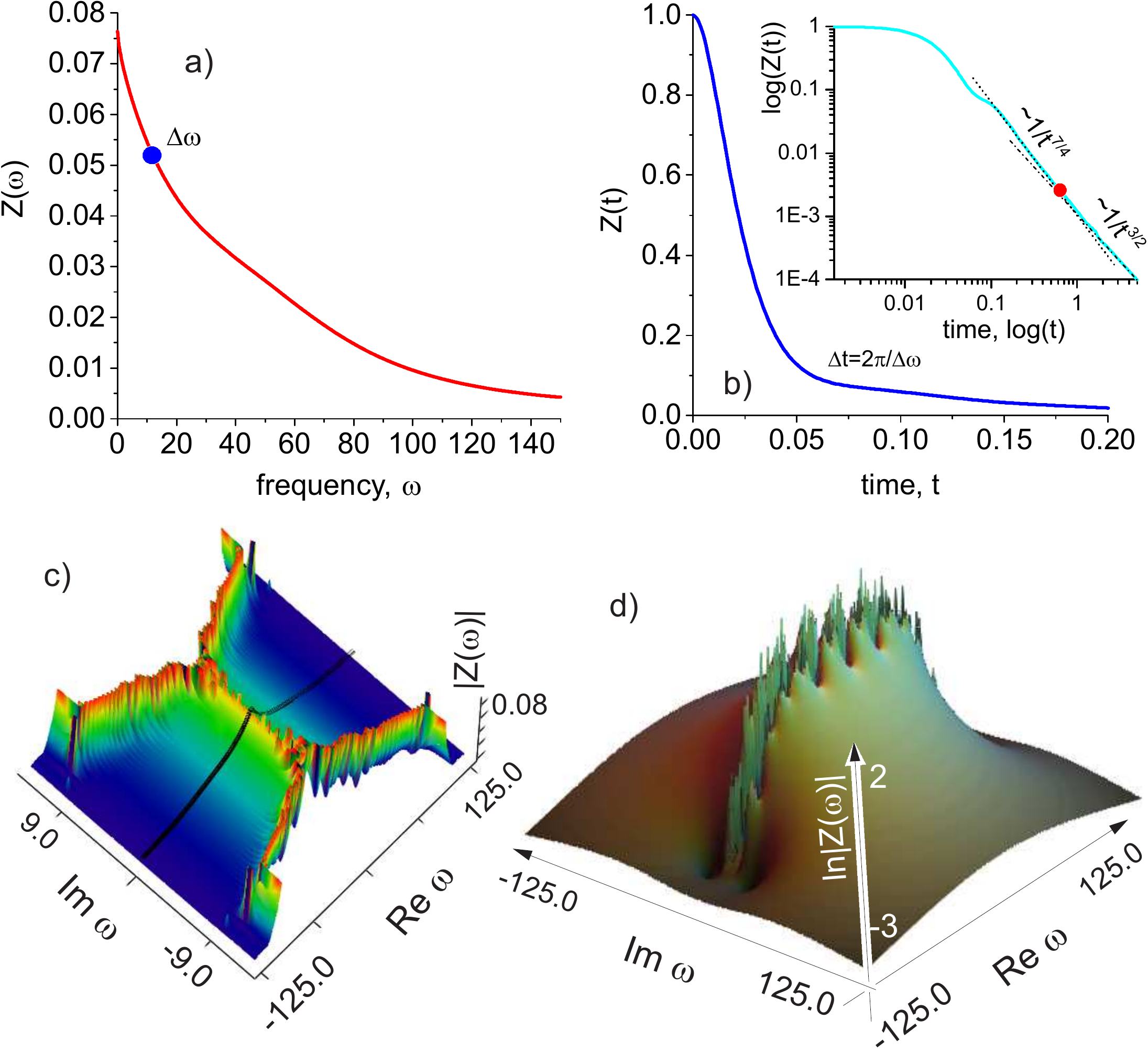}\\
  \caption{  VAF at real frequency axis (a) and in time representation (b) for $T=40$ and $\rho=1$: results of MD simulation. Insert in (b) shows $Z(t)$ in double logarithmic scale. Graphs (c) and (d) show $|Z(\omega)|$ and  $\ln|Z(\omega)|$ analytically continued into complex $\omega$-plain using multipoint Pade approximation built on top of $1400$ uniformly distributed knot-points in $\omega\in(0,200)$. The (half) width of the branch cut $\Delta \omega$ we mark by the blue bullet \textcolor{blue}{$\bullet$} on $Z(\omega)$.  Characteristic time $\Delta t=2\pi/\Delta \omega$ we show by the red bullet \textcolor{red}{$\bullet$} on $Z(t)$ curve. }\label{figT40}
\end{figure}

For density $\rho=1$ the temperature $T=40$ is characteristic temperature when fluid local structure vanishes, see Ref.~\cite{Ryltsev2013PRE}. However the branch cut in $Z(\omega)$ is still  well observable, see Fig.~\ref{figT40}. We see from Fig.~\ref{figT40}(c) that the ``small'' cut originating from $\sqrt \omega$ singularity at $\omega\to 0$ continuously transforms into the ``large'' branch cut that goes parallel to the real $\omega$-axis.

Insert in Fig.~\ref{figT40}(b) shows $Z(t)$ in double logarithmic scale. Dash-dotted line there sketches the hydrodynamic $\propto1/t^{3/2}$-tail. The red bullet shows $\Delta t$. It follows that again $\Delta t\sim t_h$.
\begin{figure}[t]
  \centering
  \includegraphics[width=0.99\columnwidth]{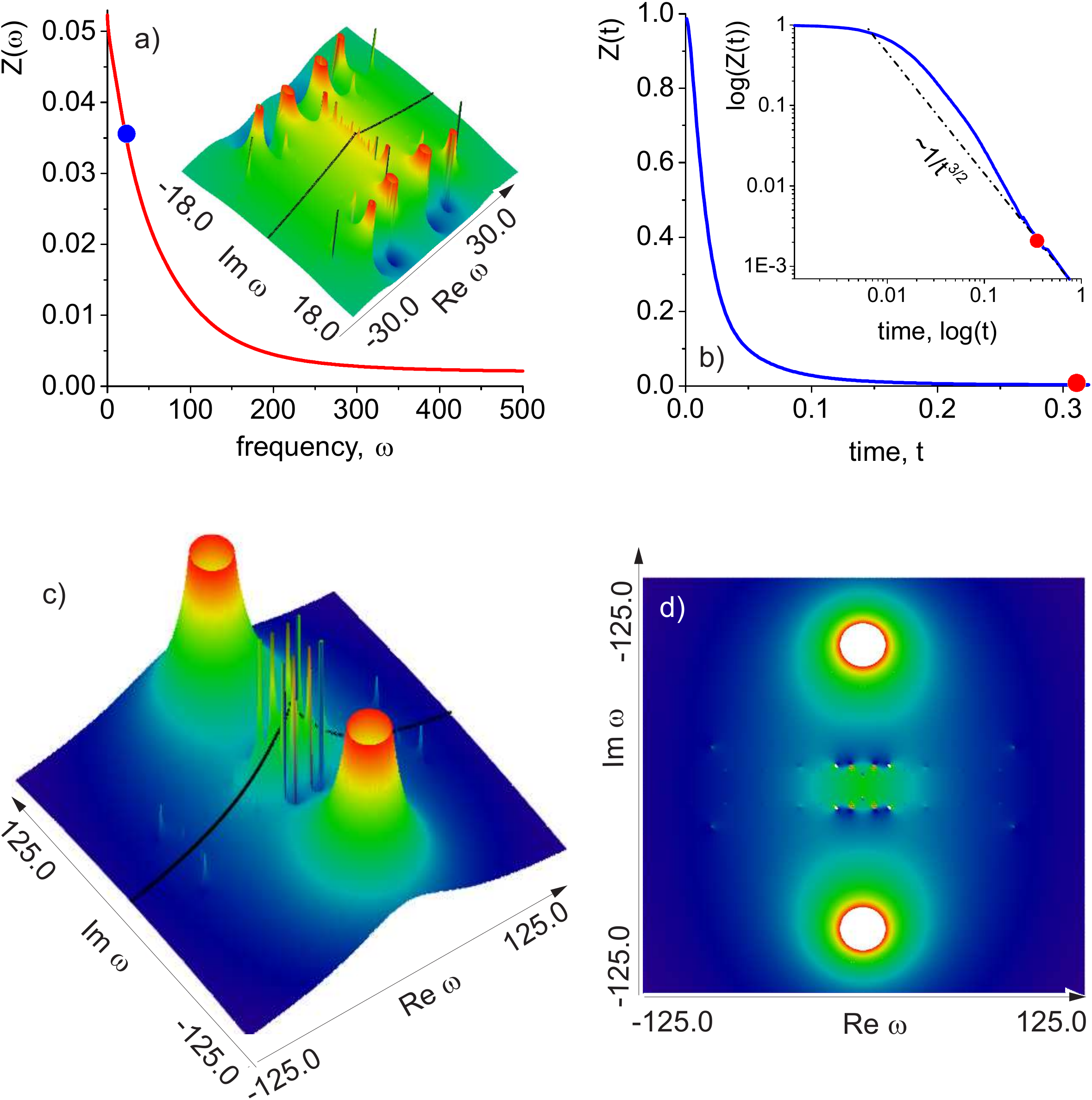}\\
  \caption{  VAF at real frequency axis (a) and in time representation (b) for $T=200$ and $\rho=1$: results of MD simulation. Graphs (c) and (d) show 3D and density plots of $|Z(\omega)|$ in complex $\omega$-plain. Multipoint Pade approximation has been built on top of $1400$ uniformly distributed knot-points in $\omega\in(0,500)$. Again, the (half) width of the branch cut $\Delta \omega$ we mark by the blue bullet \textcolor{blue}{$\bullet$} on $Z(\omega)$.  Characteristic time $\Delta t=2\pi/\Delta \omega$ we show by the red bullet \textcolor{red}{$\bullet$} on $Z(t)$ curve. }\label{figT200}
\end{figure}
\begin{figure}[b]
  \centering
  \includegraphics[width=\columnwidth]{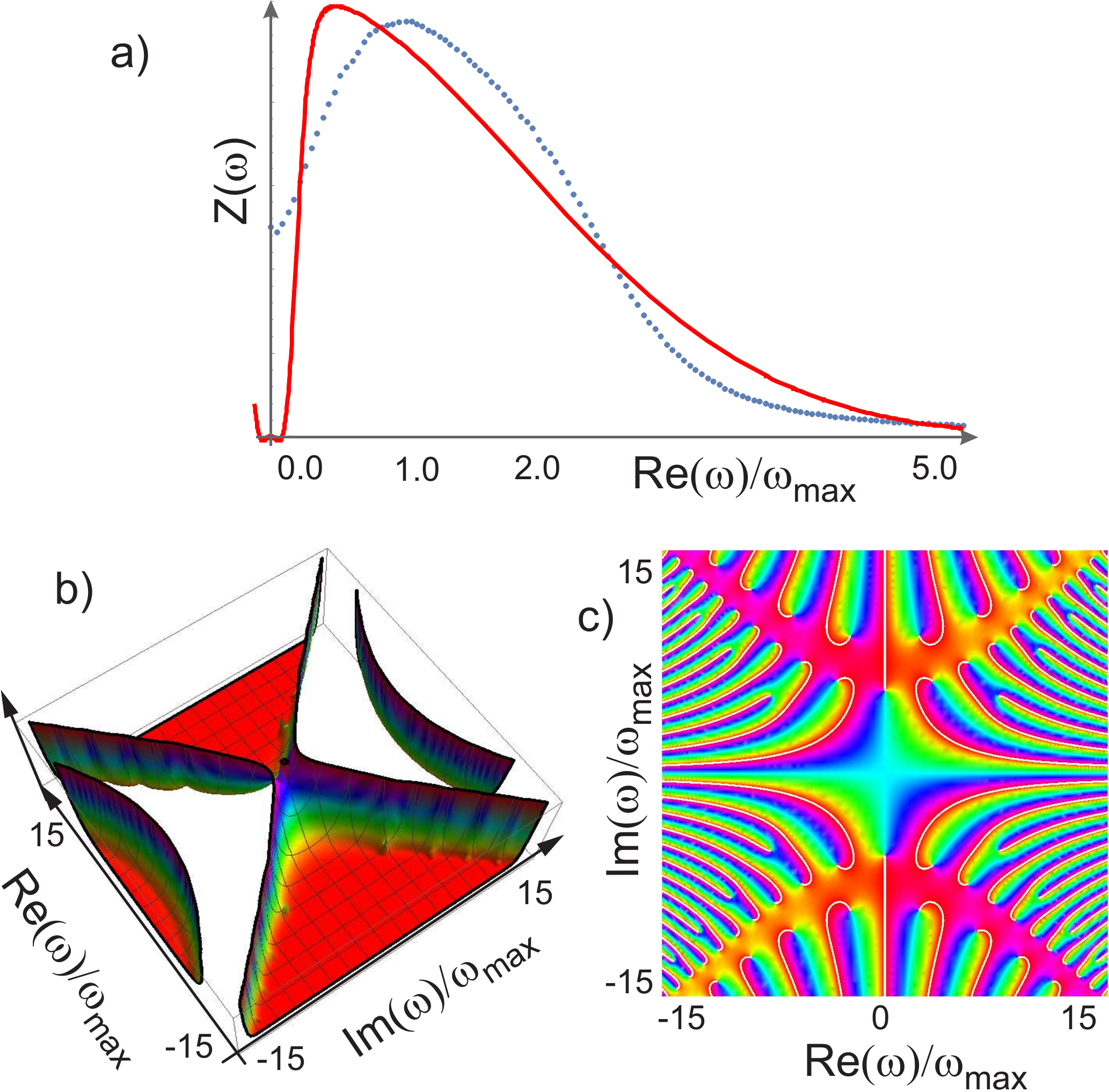}\\
  \caption{  (a) The blue dotted curve is VAF obtained by MD simulation for $T=1.4$ and $\rho=1$ while the red curve is the approximation when the memory function contains two distinct relaxation processes. Graphs (b) and (c) show the absolute value and the complex argument of the approximation analytically continued to the complex plain. } \label{fig_two_times}
\end{figure}

\subsubsection{$T=200$ and $\rho=1$}
For density $\rho=1$ and temperature $T=200$ we have the slightly nonideal gas. However even at such high temperatures there is a branch cut, very small one, as follows from Fig.~\ref{figT200}. Insert in Fig.~\ref{figT200}(b) shows $Z(t)$ in double logarithmic scale, where the  dash-dotted line sketches $\propto1/t^{3/2}$-tail. The red bullet we put at $t=\Delta t$. Again, $\Delta t\sim t_h$.

Except small branch cuts, Fig.~\ref{figT200}(d,e) reveal two isolated poles located on the imaginary axis. The appearance of such poles at hight temperatures shows that the dynamics of the system is near to that for the ideal gas. Indeed the simplest low-density-limit exponential relation for $Z(t)$ obtaining from either the Enskog approximation or the Brownian one has the same analytical structure with two pure imaginary poles~\cite{hansenMcDonald}.

\section{Discussion}

Velocity autocorrelation function in general can be expressed as follows in the Fourier space~\cite{hansenMcDonald}:
\begin{gather}%\label{}
 \tilde Z(\omega)=\frac A{-i\omega+M(\omega)},
\end{gather}
where $A$ is constant and $M(\omega)$ is the ``memory'' function. Here ``tilde'' above $Z$ means that we put $Z(t<0)=0$. If we say that $Z$ depends on $|t|$ then $Z(\omega)=(\tilde Z(\omega) +\tilde Z(-\omega))/2$.

Using the projector operator formalism \cite{hansenMcDonald,boon1991MolHydroBook} it is possible to find an exact representation of the memory function as the continued fraction of the form
 \begin{gather}\label{M_fraction}
M(\omega ) = \frac{{M_1 (0)}}{{ - i\omega  + \frac{{M_2 (0)}}{{ - i\omega  +  \ldots }}}},
\end{gather}
where $M_n(\omega)$, $n>0$ is the hierarchy of memory functions. The coefficients $M_n(0)$ are related to the frequency moments and may be in principle calculated through interaction potential and static properties. In practice the only few first moments can be calculated and so one usually has to truncate the continued fraction (\ref{M_fraction}) at some finite term~\cite{Mokshin2015TMF}. The simplest case of the first-order truncation $M_n(0)=0, n>1$ gives trivial exponential decay of VAF; the second one corresponds to non-trivial case of  $M(\omega)\propto 1/(-i\omega +1/\tau)$~\cite{hansenMcDonald} which demonstrates qualitatively correct VAF behaviour but quantitatively fails to approximates $Z(\omega)$ even at real $\omega$. As a matter of fact, no finite truncation scheme gives quantitative description at whole $\omega$ range. This leads researchers to use phenomenological approximations for memory functions, see \cite{boon1991MolHydroBook} for the review. The general conclusions about such approximations is that one relaxation time models are not enough to describe $Z(\omega)$ satisfactory; at last two relaxation times is needed. Below we consider one of such approximation as well as alternative approach based on mode coupling theory and investigate what behaviour of $Z$ in the complex $\omega$ plain these approaches produce.

It should be noted here that the time evolution of $Z(t)$ can be qualitatively understood if we truncate the continued fraction of $M$. Then  $\tilde Z(\omega)$ has few poles. When the poles of $Z(\omega)$ are purely imaginary then $Z(t)$ should decay monotonically and the poles correspond to the relaxation times; the nonzero real part of the poles induce nonmonotonic behavior of $Z(t)$~\cite{hansenMcDonald}.  Unfortunately this approximation usually has very poor accuracy~\cite{hansenMcDonald}. These considerations also fail explaining the hydrodynamic time scales, $t\gg\tau$, where $Z(t)$ shows the universal long-time tails $Z(t)$  governed by hydrodynamic fluctuations, see Refs.~\cite{hansenMcDonald,Ernst1970PRL,Dorfman1970PRL,Ernst2005PRE,Ryltsev2014JCP}.

\subsection{VAF: Two-exponential approximation of the memory function}

There is well known approximation involving two relaxation times, where~\cite{Levesque1970PhysRevA,hansenMcDonald}:
\begin{gather}%\label{}
  M(t>0)\sim A t^4e^{-a t}+B e^{-b t^2}.
\end{gather}
Taking $A,B,a$ and $b$ as an adjusting parameters and doing Fourier transform of $M(t)$ we can fit VAF. The result is illustrated in Fig.~\ref{fig_two_times}(a) for $T=1.4$ and $\rho=1$. The fit is not very good however it seems from the first glance that main features of VAF this approximation reproduces, at least at moderate and large frequencies. However going to the complex $\omega$-plain we see absolutely different behaviour than exact VAF shows, see Fig.~\ref{fig_two_times}(b) and (c): there is no branch cut parallel to the real axis. Below we investigate better approximation, however it also does not show coincidence with the exact result in the complex plain.

\begin{figure*}[t]
  \centering
  \includegraphics[width=\textwidth]{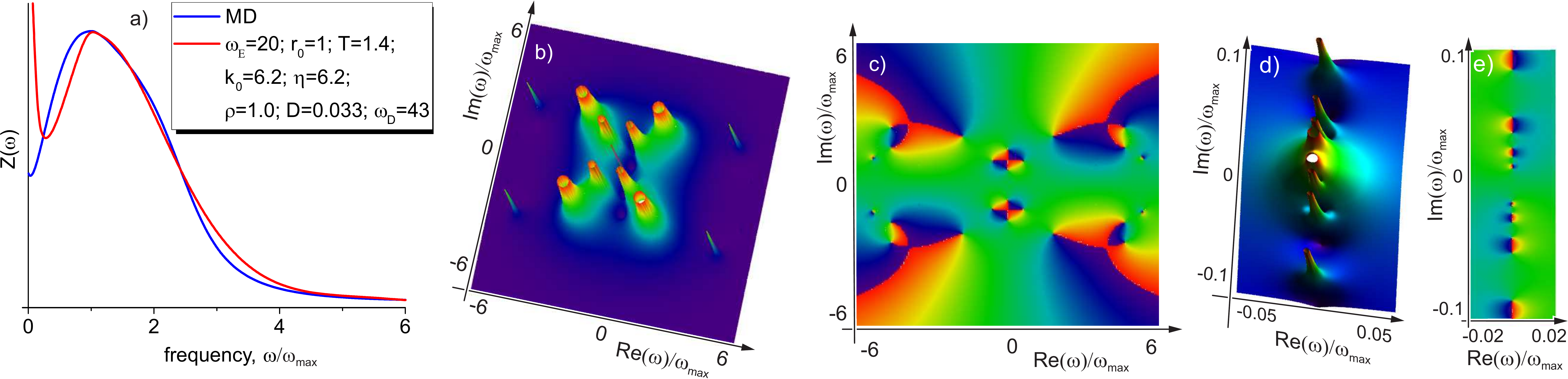}\\
  \caption{  The black curve in (a) is VAF obtained within MD simulation while the red curve in (a) shows VAF obtained within the viscoelastic model. Analytical continuation of  VAF in viscoelastic model into complex frequencies: (b) represents the amplitude of VAF while (c) is its complex phase. No brunch cut is seen. Graphs (d) and (e) show behaviour of $Z(\omega)$ at small (hydrodynamic) frequencies. 1000 uniformly distributed points of $Z(\omega)$ have been used to built the Pade approximant. Parameters: $T=1.4$ and $\rho=1$.} \label{fig_VAF_visco}
\end{figure*}
\subsection{Mode coupling approach and viscoelastic approximation}
An alternative way to calculate VAF based on mode coupling theory says that $Z(t)$ can be expressed in the form \cite{hansenMcDonald}:
\begin{gather}\label{Zvisco}
  Z(t)=\frac{3}{2\pi ^{2} \rho } \int _{0}^{\infty }f(k)\,F_{s}  (k,t)\left[C_{l} (k,t)+2C_{t} (k,t)\right]dk,
\end{gather}
where $F_s$ is the self part of intermediate scattering function, while $C_l$ and $C_t$ are the longitudinal and transverse current autocorrelation functions~\cite{hansenMcDonald}. The wight-function $f(k)$ is provides the appropriate cut-off of the integral at large enough $k$ \cite{GaskellMillerJPhys1978}.

The analytical calculation of the integrand in (\ref{Zvisco}) is the non-trivial task. The only $F_s(k,t)$ can be estimated relatively easy within the framework of the gaussian approximation: $F_s(k,t)\simeq \exp(-\alpha(t)k^2)$, where $\alpha(t)$ is an unknown function which may be related to either the mean-square displacement \cite{hansenMcDonald} or VAF itself \cite{gaskell1978}. The calculation of $C_l$, $C_t$ is a more complicated task. Here we use the simplest approximation -- viscoelastic one.

 For $C_l$ we have used the following standard expression~\cite{hansenMcDonald}:
\begin{gather}%\label{}
  C_l=\frac1\pi\frac{\omega^2 \omega_0^2 \Real N_l}{\left(-\omega \Imag N_l-\frac{\omega_0^2}{S(k)}+\omega^2\right)^2+(\omega  \Real N_l)^2},
  \\
  N_l(\omega,k)=\frac{\omega_l^2(k)-\frac{\omega_0^2(k)}{ S(k)}}{-i\omega+\gamma_l(k)}.
\end{gather}
Here $\omega_0=k\sqrt{T/m}$ and $S(k)$ is the structure factor that we take from MD simulation. The effective frequencies are defined as follows~\cite{GaskellMillerJPhys1978}:
\begin{multline}%\label{}
 \omega_l^2(k)=3 \omega_0^2(k)+
 \\
 \omega_E^2 \left(-\frac{3 \sin (k r_0)}{k r_0}+\frac{6 \sin (k r_0)}{(k r_0)^3}-\frac{6 \cos (k r_0)}{(k r_0)^2}+1\right),
\end{multline}
where $\omega_E$ and $r_0$ are the adjusting parameters of the order of the Einstein frequency and interparticle spacing. The damping~\cite{GaskellMillerJPhys1978,hansenMcDonald}
\begin{gather}%\label{}
 \gamma_l^2(k)=\frac 4\pi\left\{\omega_l^2(k)-\frac{\omega_0^2(k)}{S(k)}\right\}.
\end{gather}

For $C_t$ the following expression have been used~\cite{GaskellMillerJPhys1978,hansenMcDonald}:
\begin{gather}%\label{}
C_t=\frac{k^2 T G(k)}{\rho \tau_t(k) \left(\left(\omega^2-\frac{k^2 G(k)}{\rho }\right)^2+\left(\frac{\omega }{\tau_t(k)}\right)^2\right)},
\end{gather}
where
\begin{gather}%\label{}
  \tau^{-1}_t(k)=\frac{\left({G_0/\eta }\right)^2-{2 k^2 (G(k)/\rho-T)}}{k^2r_0^2+1}+\frac{2 k^2 G(k)}{\rho },
  \\
  G_0=\frac{1}{10} \rho  r_0^2\omega_E^2+\rho  T,
  \\
  G(k)=\frac{\rho  \left(\omega_0^2(k)+\omega_E^2 \,\left(-\frac{3 \sin (k r_0)}{(k r_0)^3}+\frac{3 \cos (k r_0)}{(k r_0)^2}+1\right)\right)}{k^2},
\end{gather}
where $\eta$ is viscosity (we take it from MD simulations).

For the dynamic structure factor we have used the approximation following Refs.~\cite{egelstaff1962,Copley1975JProgPhys}:
\begin{gather}%\label{}
  F_{s} (k,t)\approx \exp \left[-Dk^{2} \left(\sqrt{t^{2} +c^{2} } -c\right)\right],
\end{gather}
where $c\approx1/\omega _{E} $ and $D$ is the diffusion coefficient (we take it from MD simulations).  This approximation has perfect analytical form in the Fourier space:
\begin{multline}
S_{s} (k,\omega )=\frac{1}{\pi } \frac{Dk^{2} /\omega _{D} }{\sqrt{\omega ^{2} +(Dk^{2} )^{2} } }\times
\\
\exp \left(\frac{Dk^{2} }{\omega _{D} } \right)K_{1} \left[\frac{\sqrt{\omega ^{2} +(Dk^{2} )^{2} } }{\omega _{D} } \right].
\end{multline}
Here $K_{1}$ is the bessel function.

For $\rho=1$ and $T=1.4$ (LJ fluid) we find the best fit of $Z(\omega)$ using the Viscoelastic approximation.   The results are shown in Fig.~\ref{fig_VAF_visco}. The best fit parameters used in Fig.~\ref{fig_VAF_visco} are in fact very close to those estimated independently from molecular dynamic simulations. The black curve in (a) is VAF obtained within MD simulation while the red curve in (a) shows VAF obtained within the viscoelastic model. Analytical continuation of  VAF in viscoelastic model into complex frequencies is shown in (b), (c), (d) and (e). As follows from (a) the difference between viscoelastic and exact results for $Z(t)$ is not very large however no brunch cut is seen in viscoelastic $Z(t)$, see (b) and (c), contrary to exact $Z(t)$. Similar result we see for for other densities and temperatures for LJ fluid.

Analytical continuation given in Fig.~\ref{fig_VAF_visco} shows that $Z(\omega)$ defined by a quite involved integral in the viscoelastic model can be unexpectedly well approximated just by the analytical function with the poles at $\omega_1=\pm 0.57\pm i 0.97 $ and $\omega_2=1.74+i 1.87$:
\begin{gather}\label{eq:ZL}
  \tilde Z(\omega)\approx \sum_{i=1,2}\frac{ A_i}{-i\omega+\frac{|\omega_i|^2}{-i\omega+2\omega_i''}},
\end{gather}
where $A_i$ are the real adjusting parameters. Only at very small (hydrodynamic) frequencies this simple approximation becomes incorrect. It should be also noted that the longitudinal part of current fluctuations, see Eq.~\eqref{Zvisco}, gives the main contribution to $Z(\omega)$ except the peak at small frequencies where transverse fluctuations dominate.

Summarizing this section, one can again conclude that there is no analytical approach to calculate VAF with enough accuracy to build analytical continuation in complex frequency plain. So the only alternative is the numerical methods described below.

\section{Methods~\label{secMethods}}
\begin{figure}[t]
  \centering
  \includegraphics[width=\columnwidth]{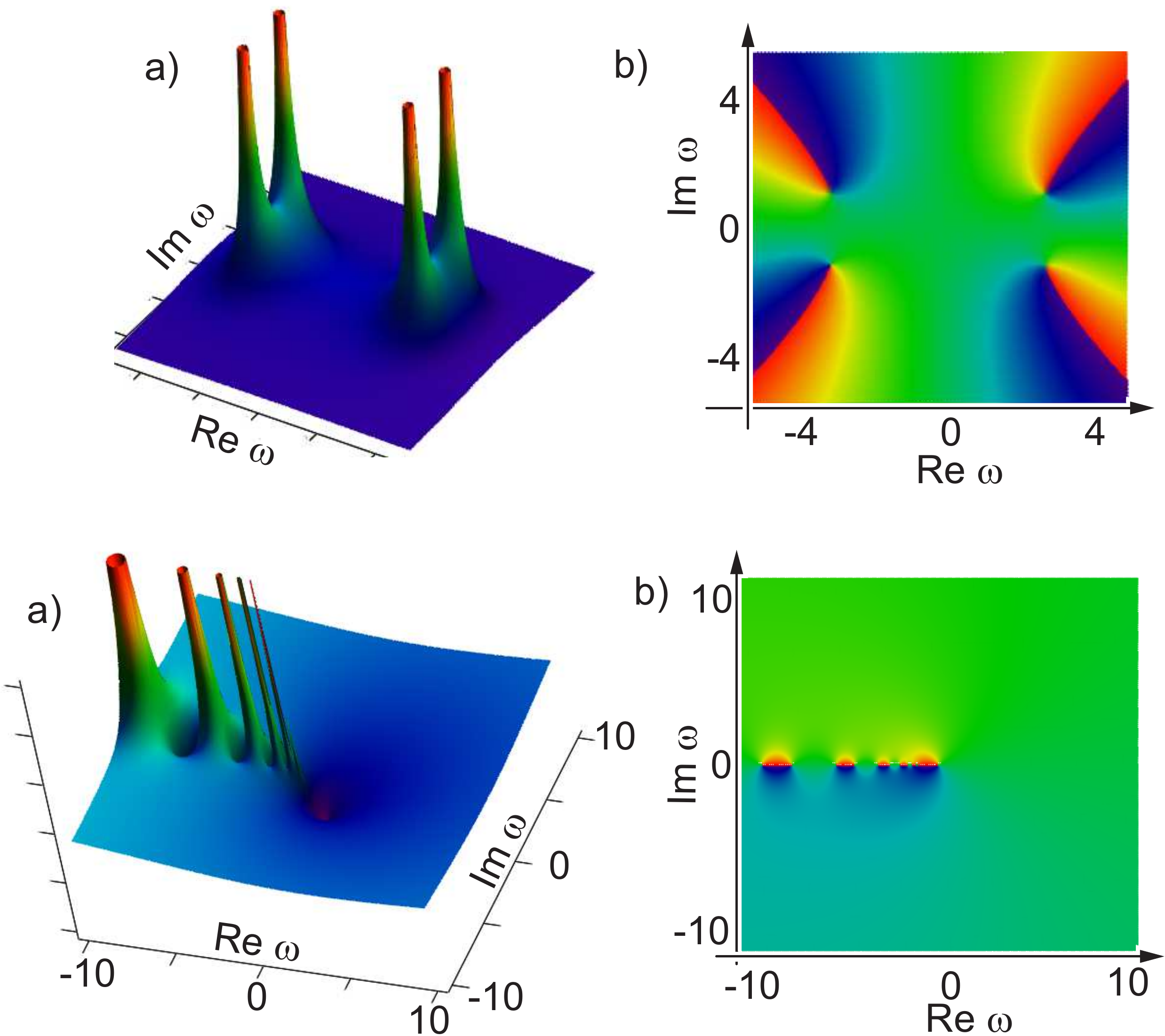}\\
  \caption{  Analytical continuation of oscillator power spectrum from the real axis to the complex plain by multipoint Pade approximant: (a) shows the absolute value while (b) is the argument. } \label{fig_ap_osc}
\end{figure}

\subsection{Pade approximation: Numerical multipont  continued fraction algorithm}

Here we discuss the construction of the Pad\'{e} approximants that interpolate a function given $N$ knot points.  Pade-approximants are the rational functions  (ratio of two polinomials).   A rational function can be represented by a continued fraction. Typically the continued fraction expansion for a given function approximates the function better than its series expansion.

Algorithm: for a function $f(x_i)=u_i$ with values $u_i$ at $N$ knots $x_i$, $i=1,2,3,\ldots,N$, the Pade approximant is
\begin{gather}%\label{}
C_N(x)=\frac{a_1}{\frac{a_2\left(x-x_1\right)}{\frac{a_3 \left(x-x_2\right)}{\frac{a_4\left(x-x_3\right)}{\ldots  a_N\left(x-x_{N-1}\right)+1}+1}+1}+1}
\end{gather}
where  $a_i$ we determine using the condition, $C_N(x_i)=u_i$, which is fulfilled if $a_i$ satisfy the recursion relation
\begin{gather}%\label{}
  a_i=g_i\left(x_i\right), \qquad
g_1\left(x_i\right)=u_i, \qquad i=1,2,3,\ldots ,N.
\\
g_p(x)=\frac{g_{p-1} (x_{p-1})-g_{p-1}(x)}{\left(x-x_{p-1}\right) g_{p-1}(x)},\qquad p\geq 2.
\end{gather}

\subsection{Pade approximation: Illustrating test-examples}
\subsubsection{Oscillator power spectrum amplitude: approximation of the analytical function with 4 poles.}
\begin{figure}[t]
  \centering
  \includegraphics[width=\columnwidth]{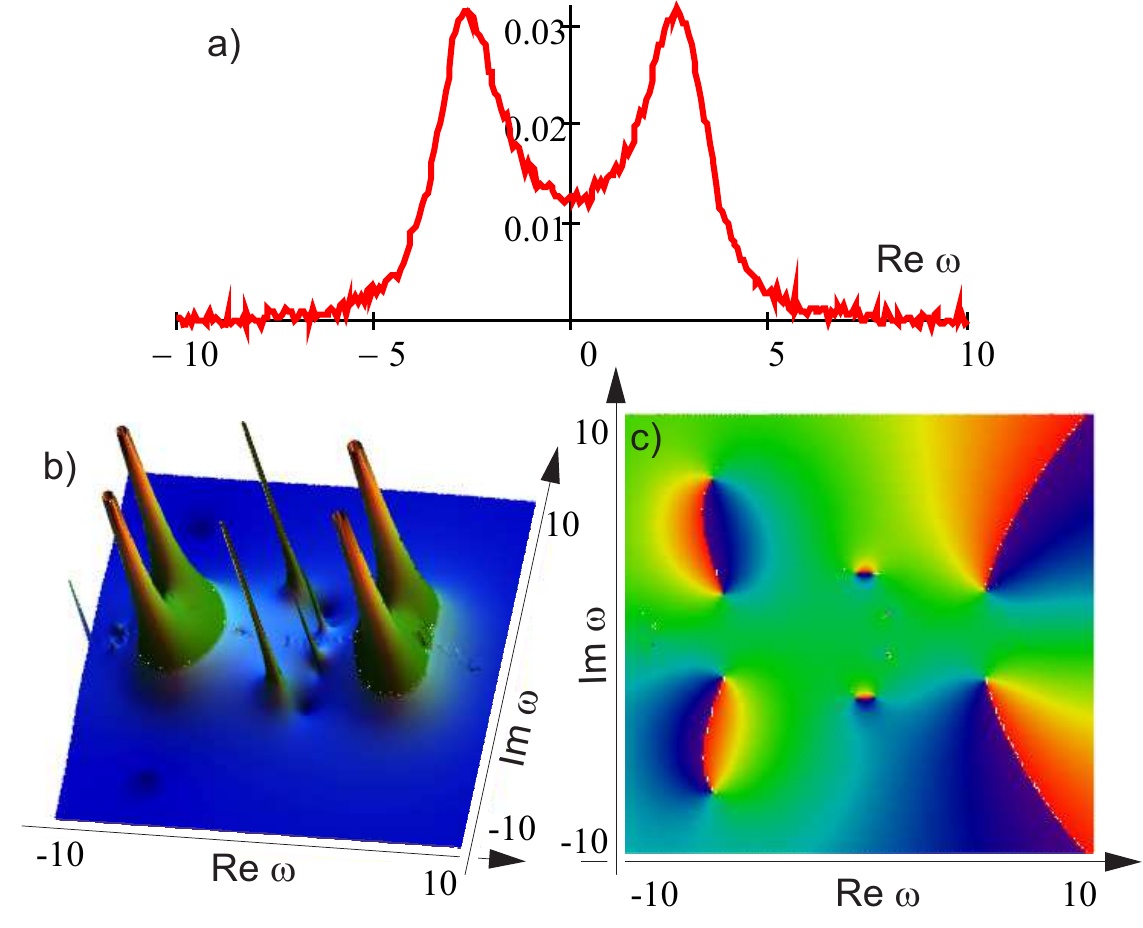}\\
  \caption{  (a) We add gaussian noise with zero mean and $\sigma=5\times10^{-4}$ to the oscillator power spectrum.  Analytical continuation is shown in (b) and (c). The ``main'' poles show high degree of resistivity to noise.} \label{fig_ap_osc_noise}
\end{figure}
\begin{figure}[b]
  \centering
  \includegraphics[width=\columnwidth]{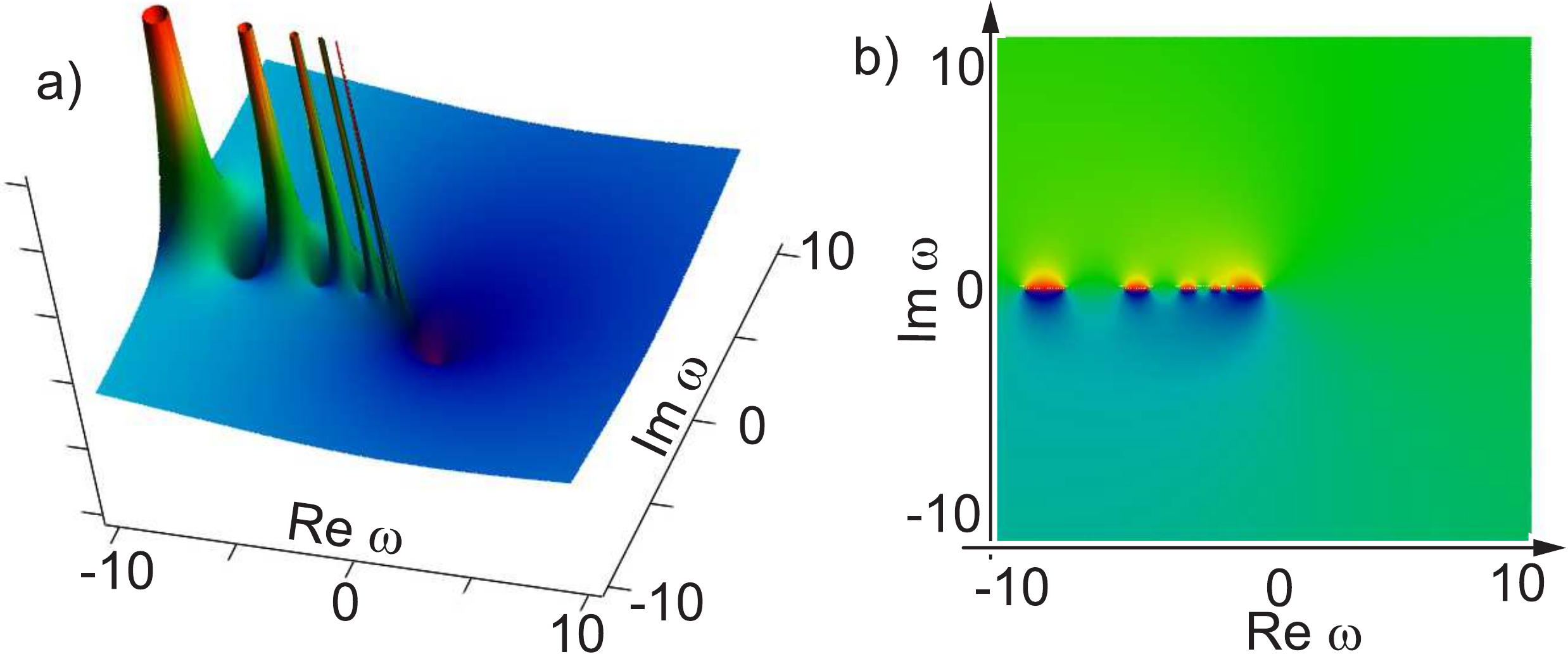}\\
  \caption{  Analytical continuation of the logarithm: (a) shows the absolute value while (b) is the argument. } \label{fig_ln}
\end{figure}

\begin{figure}[t]
  \centering
  \includegraphics[width=\columnwidth]{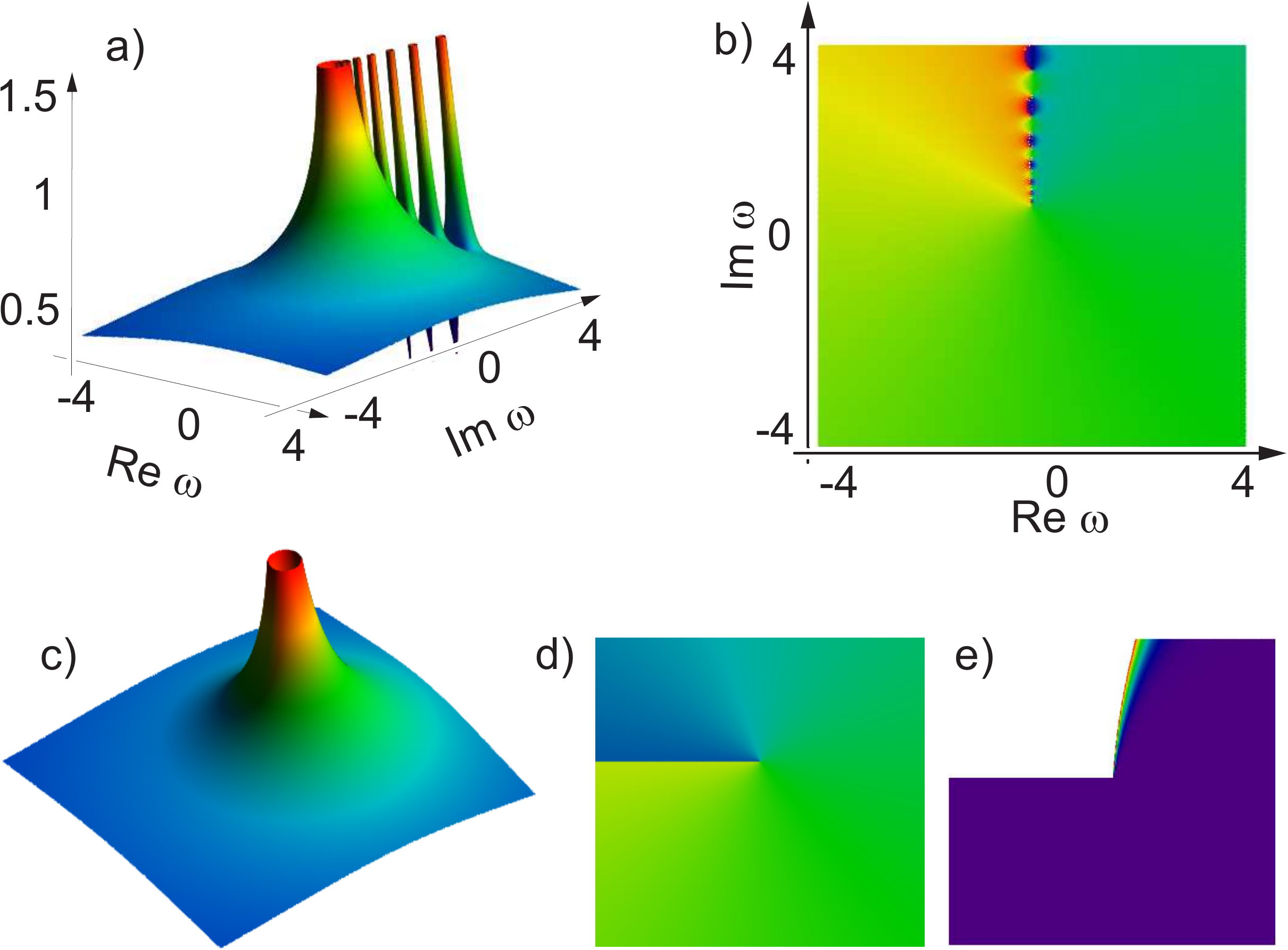}\\
  \caption{  Analytical continuation of the square root function: (a) shows the absolute value  while (b) is the argument. Array of peaks and dips in (a) represent the branch cut. Graphs (c) and (d) show ``exact'' absolute value and argument of the function. Figure (e) is the density plot of the relative difference between the exact and Pade-approximation. The coincidence is perfect everywhere except the white zone where the functions differ because the branch cuts of the approximation and the ``exact function'' have been chosen differently.} \label{fig_sqrt_pole}
\end{figure}

As the first test example we take the function
\begin{gather}\label{fosc}
 f_{\rm osc}(\omega)=\frac 1{(\omega^2-\omega_0^2)^2+(\omega \gamma)^2}.
\end{gather}
This function is proportional to the oscillator power spectrum. We take $\omega=3$ and $\gamma=2$ and build the Pade approximant using 300 uniformly distributed knots at $\omega\in(-10,10)$. The result of the analytical continuation is show in Fig.~\ref{fig_ap_osc}. We worked with double precision. The relative error of the analytical approximation was less than $10^{-10}$ even in the pole-regions.

\subsubsection{Stability of the Pade-approximation}
We add gaussian noise with zero mean and $\sigma=5\times10^{-4}$ to the oscillator power spectrum considered above, see Fig.~\ref{fig_ap_osc_noise}.  Analytical continuation is shown in (b) and (c). The ``main'' poles are still clearly seen. So analytical continuation by Pade approximation is quite resistive to noise if the noise correlation length is short enough.

\subsubsection{Analytical continuation of $ln$-function by the Pade approximant}
Now we illustrate how behaves singular function in the complex plain when we do its Pade analytical continuation. We take
\begin{gather}\label{fln}
 f_{\rm ln}(\omega)=\ln(1+\omega).
\end{gather}
We build the Pade approximant using 300 uniformly distributed knots at real $\omega\in(0,50)$.  The result of the analytical continuation is shown in Fig.~\ref{fig_ln}. There is cut $(-\infty,-1)$ in the complex $\omega$-plain. Analytical continuation based on the Pade approximant  reproduces the cut by the array of poles and knots (where $ f_{\rm ln}=0$), see Fig.~\ref{fig_ln}. Away from the cut the accuracy of the analytical continuation is satisfactory as in the upper illustrating example while $|\omega|<50$.

\begin{figure}[t]
  \centering
  \includegraphics[width=\columnwidth]{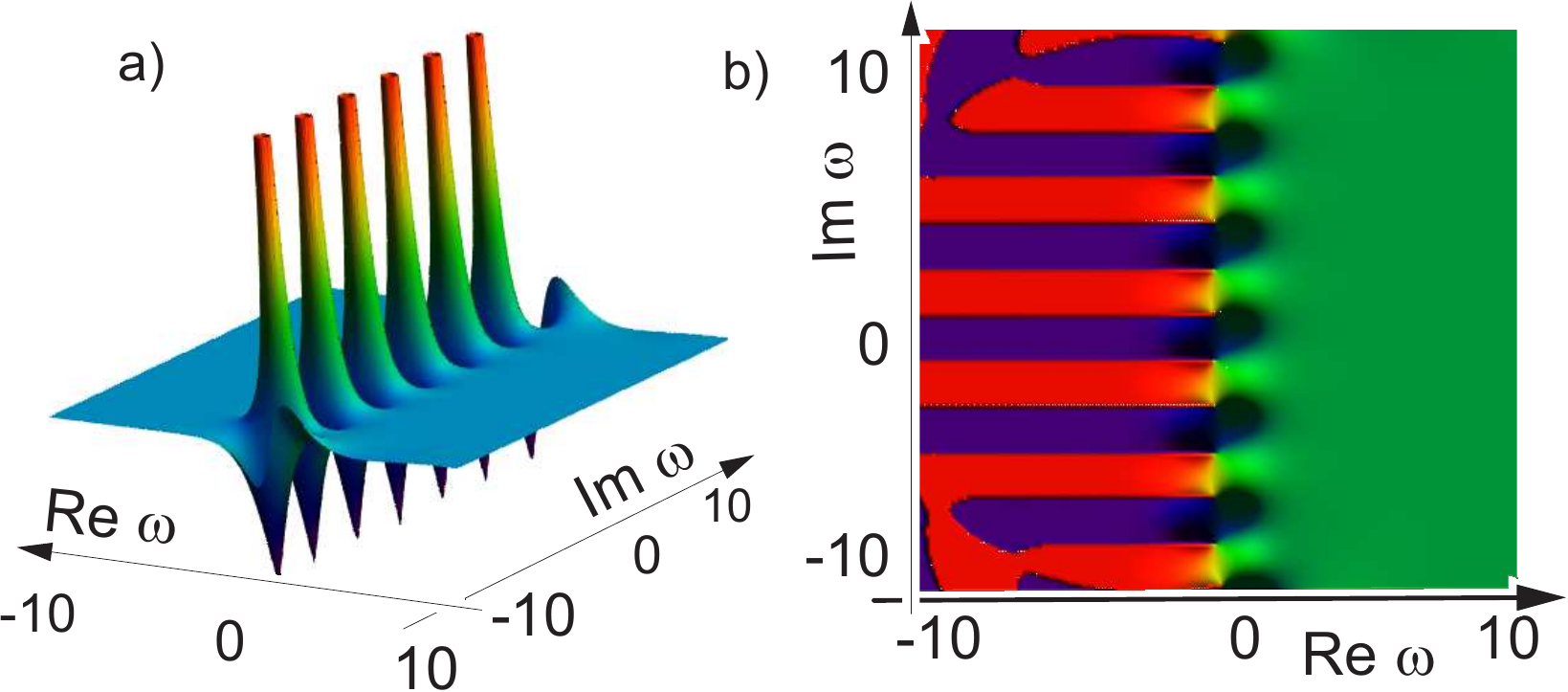}\\
  \caption{  Analytical continuation of $\tanh(\omega)$: (a) shows the absolute value  while (b) is the argument. We build the Pade approximant using 500 uniformly distributed knots at $\omega\in(-10,10)$. } \label{fig_tanh}
\end{figure}
\subsubsection{Analytical continuation by the Pade approximant of the function with  the square root singularity. }
Finally we take the function with  the square root singularity to test the Pade-approximation:
\begin{gather}\label{fsqrt}
 f_{\rm sqr}(\omega)=\sqrt{\omega-i}.
\end{gather}
We build the Pade approximant using 300 uniformly distributed knots at real $\omega\in(-10,10)$. Then we analytically continue the Pade polinomial (it is in fact complex even at real $\omega$) to the complex $\omega$-plane as shown in Figs.~\ref{fig_sqrt_pole}(a) and (b).  Array of peaks and dips in (a) represent the branch cut: this is typical for pade approximation. Graphs (c) and (d) show ``exact'' absolute value and argument of the function. The branch cut parallel to the real axis is typical choice for ``computer'' build in functions (we have used Mathcad). Pade approximation have chosen different direction for the branch cut, parallel to the imaginary axis, see (a) and (b). Figure (e) is the density plot of the absolute value of the difference between the exact and Pade-approximation ($|f_{\rm ln}-f_{\rm ln}^{\rm( pade)}|/(|f_{\rm ln}|-|f_{\rm ln}^{\rm( pade)}|)$. The coincidence is perfect everywhere except the white zone where the functions differ because the branch cuts of the Pade approximation and the ``exact function'' are different.

\subsection{Limits of applicability of Pade approximation}
As follows from the examples, if we approximate a function by the Pade polinomial at certain domain at the real axis then the analytical continuation is more or less perfect at the circle in the complex plain (around that domain) with the radius about the length of the domain.

The branch cuts are represented by an array of poles.

There is a problem with the branch cuts: we can draw them differently in the complex plain, only edges are fixed. Different choice of the branch cut curve corresponds different analytical continuation. But the Pade polynomial chooses the cut curve somehow ``automatically'': we do not well control that.  So the Pade approximation is a useful tool if one needs to identify the position and types of the singularities of the function in the complex plain like poles and the the branch cut edges. For functions without branches analytical continuation in unique and the Pade approximation well produces it, see, e. g., Fig.~\ref{fig_tanh} .

\section{Conclussions}
Singularities of dynamic correlation functions in complex plane usually correspond to the
collective or localized modes.  We have found
that instead of few number of isolated poles velocity autocorrelation function of LJ particle
system in complex plain shows the singularity manifold forming branch cuts that suggests LJ velocity autocorrelation function is a
multiple-valued function of complex frequency. The brunch cuts are separated from the real axes
by the well-defined ``gap''.  The brunch cuts are quite stable with the respect to temperature and density variation. We have found the trace of the singularity gap at temperatures several orders of magnitude higher than the melting temperature.  Our working hypothesis is that the branch cut origin is related to the interference of short-time one-particle kinetics and and long-time collective motion (hydrodynamics).

\acknowledgments

This work was supported by Russian Scientific Foundation (grant RNF №14-12-01185). We are grateful to Russian Academy of Sciences for the access to JSCC and ``Uran'' clusters and National Research Centere ``Kurchatov Institute'' for access to HCP-supercomputer cluster.

 \bibliography{our_bib}
\end{document}